\documentclass[%
 reprint,onecolumn,
 amsmath,amssymb,
 aip, 12pt
]{revtex4-2}
\usepackage{xcolor}
\usepackage{graphicx}% Include figure files
\usepackage{dcolumn}% Align table columns on decimal point
\usepackage{bm}% bold math
\usepackage{hyperref}
\usepackage{setspace}
\usepackage{threeparttablex}
\usepackage{booktabs}
\usepackage{array}
\usepackage[title]{appendix}
\newcommand{\bra}[1]{\ensuremath{\left\langle#1\right|}}
\newcommand{\ket}[1]{\ensuremath{\left|#1\right\rangle}}
\newcommand{\ketbra}[2]{\ensuremath{\left|#1\right\rangle\left\langle#2\right|}}

\newcolumntype{P}[1]{>{\centering\arraybackslash}p{#1}}
\begin{document}

\title{On the nature of polariton transport in a Fabry-Perot Cavity}
\author{Zeyu Zhou}
\affiliation{Department of Chemistry, University of Pennsylvania, 231 South 34th Street, Philadelphia, Pennsylvania 19104, United States}
\author{Hsing-Ta Chen}
\affiliation{Department of Chemistry, University of Pennsylvania, 231 South 34th Street, Philadelphia, Pennsylvania 19104, United States}
\affiliation{Department of Chemistry and Biochemistry, 251 Nieuwland Science Hall, Notre Dame, Indiana 46556, United States}
\author{Maxim Sukharev}
\affiliation{Department of Physics, Arizona State University, Tempe, Arizona 85287, United States}
\affiliation{College of Integrative Sciences and Arts, Arizona State University, Mesa, Arizona 85212, United States}
\author{Joseph E. Subotnik}
\affiliation{Department of Chemistry, University of Pennsylvania, 231 South 34th Street, Philadelphia, Pennsylvania 19104, United States}
\author{Abraham Nitzan}
\affiliation{Department of Chemistry, University of Pennsylvania, 231 South 34th Street, Philadelphia, Pennsylvania 19104, United States}
\affiliation{Department of Physical Chemistry, School of Chemistry, The Raymond and Beverly Sackler Faculty of Exact Sciences and The Sackler Center for computational Molecular and Materials Science, Tel Aviv University, Tel Aviv 6997801, Israel}

\date{\today}

\begin{abstract}
Fabry-Pérot microcavities can strongly enhance interactions between light and molecules, leading to the formation of hybrid light-matter states known as polaritons. Polaritons possess much smaller effective masses and much larger group velocities when the molecules are resonant with cavity modes that have finite (non-zero) in-plane wavevectors, giving rise to the possibilities of long-range and ultrafast ballistic transport. 
In this paper, we present results of numerical simulations of the ultrafast ballistic transport phenomenon in real space and time during and after initialization with a short, spatially localized pulse. 
We find that the transport of the molecular excitons as induced by  the external light field is synchronized with the evolution of the enhanced and localized electromagnetic field inside the cavity. 
Moreover, the synchronized transport rate is in good agreement with the group velocities predicted from a calculated dispersion relation across a wide range of frequencies.
These simulations provide an intuitive tool for understanding the collective motion of light and excitons and helps to better understand how experimental observations of polaritons should be interpreted. 
\end{abstract}
\maketitle

\newpage

\section{Introduction}
Understanding energy transfer in materials is an active and important subject, ranging from photochemistry to solar energy harvesting. 
Because of the defects and disorder in materials that arise at finite temperatures, long-range ballistic energy transport is often suppressed and transport becomes a short-range diffusive process.
Recently, however, with advances in microcavity engineering and the introduction of high quality one and two-dimensional photonic crystals, the inherent light-matter coupling in materials can be drastically enhanced, leading to hybrid light-matter excitations and the formation of long-lived polaritons.\cite{freixanet2000propagation, bayer2001coupling,  walther2006cavity, hutchison2012modifying, feist2015extraordinary, schachenmayer2015cavity, ebbesen2016hybrid, herrera2016cavity, kavokin2017microcavities, ribeiro2018polariton, frisk2019ultrastrong, hertzog2019strong,  garcia2021manipulating, chavez2021disorder} Experiments have shown that many material properties can be modified by the formation of polaritons. Furthermore, since polaritons contain photonic components, they are able to accelerate energy transfer and extend the length scale of the ballistic energy transport.\cite{orgiu2015conductivity, lerario2017high, rozenman2018long, zakharko2018radiative, nagarajan2020conductivity, hou2020ultralong, tichauer2021multi, pandya2022tuning, sokolovskii2022enhanced, xu2022ultrafast, balasubrahmaniyam2023enhanced} 

To numerically investigate polaritonic motion, one can envision different approaches with different levels of simulation details. One approach is based on a model Hamiltonian comprising a limited number of molecular and optical degrees of freedom that is assumed to represent the hybrid light-matter system and numerically integrate the equations of motion associated with this Hamiltonian, so as to capture the ultrafast transport rate of polaritons along the in-plane direction of the cavity\cite{feist2015extraordinary, chavez2021disorder, xu2022ultrafast, tichauer2021multi, sokolovskii2022enhanced}.
One caveat of this approach is the choice of initial conditions (which in turn determines the subsequent dynamics) is not obvious.
In Ref. \citenum{xu2022ultrafast}, the molecules were modeled as 2-level emitters with initial conditions chosen using Monte Carlo sampling to rotate a set of cavity QED eigenstates (within a certain energy window) in order to minimize the final standard deviation in real space. In so doing, the authors were able to construct a non-equilibrium localized initial wavefunction that was transported in time.  

In Ref \citenum{sokolovskii2022enhanced}, an atomistic molecular model was used and, again, the electromagnetic (EM) field was represented using just a few modes. The simulations were initialized by constructing a localized gaussian of matter excitations and propagation was done using semiclassical Ehrenfest dynamics to follow the motion of the wavefront of molecular electronic excitations. In both calculations, the focus is more on molecular excitation than on the radiation field. A different approach combines a numerical solution of the Maxwell equations for the classical electromagnetic field on a real space grid with a quantum, mean-field description of the matter part. It has been demonstrated that with this level of calculation, one can capture a wide range of interesting cavity effects and important electromagnetic observables in experiments.\cite{taflove2005computational, teixeira2007fdtd, mcmahon2007tailoring, zhao2008methods, puthumpally2014dipole, sukharev2017optics, you2019nonlinear, sidler2020polaritonic, tancogne2020octopus, sukharev2021second}

In this work, we follow the latter approach. 
We will perform mixed quantum-classical calculations on a 2-dimensional grid integrating the Maxwell-Bloch equations using the finite-difference time-domain (FDTD) methodology \cite{taflove2005computational} and capture the responses of the system to an actual spatially-focused long pulse. 
Our choice of the system is a Fabry-Pérot cavity constructed using two dielectric Bragg reflectors made of two types of dielectric materials. The bare cavity yields a series of transverse magnetic (TM) modes for different incident angles $\theta$ (or equivalently, in-plane wavevectors $k_{\parallel}$) and the dispersion relation between $\omega$ and $k_{\parallel}$ for these cavity modes can be directly extracted from the FDTD calculations as well as the standard transfer matrix method (TMM).\cite{yeh1990optical, kavokin2017microcavities}
For the polariton propagation simulations, we place a single layer of molecules near the center of the cavity and use an external focused long pulse to initiate polariton excitation. 
Subsequently,  we monitor the evolution of the excited state population and extract a ballistic transport rate for each choice of incident frequency of the exciting pulse. 
We find that the calculated ballistic transport rates are in good agreement with the group velocities obtained from the dispersion relation of polaritons, and our simulations provide a new interpretation of that transport.

This manuscript is arranged as follows. 
In section \ref{sec: two}, we present our system composed of one layer of 3-level molecules and a cavity formed by two dielectric Bragg reflectors under investigation.  
In section \ref{sec: three}, we first present the dispersion relation obtained by the transfer matrix method and Maxwell-Bloch calculations and demonstrate that they agree very well. Second, we investigate the molecular excited-states population dynamics for these systems following incident spatially localized pulse excitation using numerical solution of the corresponding Maxwell-Bloch calculations. Third, we extract the transport rates obtained from Maxwell-Bloch calculations and show that they recover the group velocities predicted from the dispersion relation.
In section \ref{sec: discussion}, we analyze the underlying meaning of the group velocities and the Hopfield coefficients.
In section \ref{sec: four}, we conclude and discuss possible future directions.

\section{Model and Methods \label{sec: two}}
\subsection{System geometry}
\begin{figure}[!ht]
    \includegraphics[width=14cm]{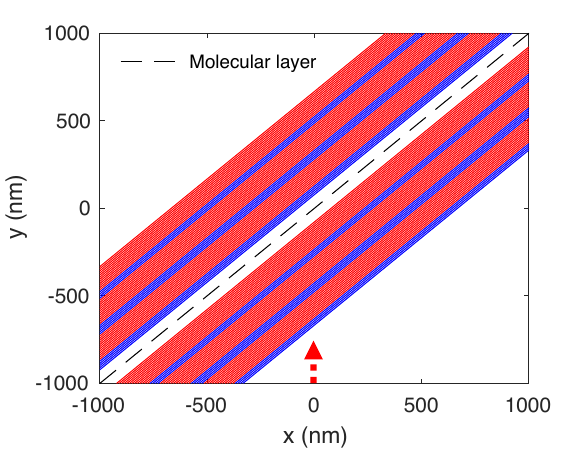}
    \caption{A schematic setup of the 2-dimensional system under investigation. The actual simulation box is $16\times 16\ \mu m^2$. A focused light source (a red dotted arrow) is placed near the edge of the simulation box and incident from the bottom of the figure. The black dashed line along the diagonal of the box represents the very thin molecular layer ($40/\sqrt{2}$ nm thickness) and therefore, we may assume that the spatial variation of the field in the direction normal to the cavity axis may be disregarded. The red and blue layers represent two distributed Bragg reflectors (DBRs) that compose the cavity. The white regions represent vacuum.}
    \label{fig: systemsketch}
\end{figure}

In this work, we consider the two dimensional system shown in  Fig \ref{fig: systemsketch}. Here, a cavity made of two distributed Bragg reflectors (each represented by $3$ pairs of red and blue stripes) is placed along the diagonal of the simulation box. This orientation ($45$ degrees with respect to the $x$ or $y$ axis) is chosen for optimal grid size, as the effective grid size is $dl = \sqrt{2}dx$. 
Both types of layers have standard quarter-wave thicknesses (depending on their refractive indices $n_{1}$ and $n_{2}$, respectively) 
according to the central energy wavelength (chosen to be $2$ eV, $\lambda_{c} = 620$ nm)
\begin{align}
    L_{1, 2} = \frac{\lambda_{c}}{4n_{1, 2}} 
    \label{eq: layerthicknesses}
\end{align}
to ensure maximal transmission for a normal incident cavity mode. By this two-dimensional setup, we focus on the transverse magnetic (TM) modes of the cavity.\cite{yeh1990optical, kavokin2017microcavities} 

A single molecular layer (black dashed line) is placed at the center of the cavity with thickness $40/\sqrt{2}$ nm. Because this layer is so thin,
the long-wavelength approximation holds and the layer will collectively respond to the external light source along the $y$ direction. 

\subsection{Maxwell-Bloch equations\label{subsec: maxwellequations}}
The symmetry of the simulated system indicates that meaningful results can be obtained using a 2-dimensional calculation.
The optical field is modeled using the three variables ${\cal E}_{x}, {\cal E}_{y}$ and ${\cal B}_{z}$, and the FDTD solver is applied to the relevant EM equations:
\begin{align}
    \frac{\partial {\cal B}_{z}}{\partial t} &= \frac{\partial {\cal E}_{x}}{\partial{y}} - \frac{\partial {\cal E}_{y}}{\partial{x}}\label{eq: mag}\\
    \epsilon\frac{\partial {\cal E}_{x}}{\partial t} &= \frac{\partial {\cal B}_{z}}{\mu_{0}\partial{y}} - J_{x}\label{eq: elecx}
    \\
    \epsilon\frac{\partial {\cal E}_{y}}{\partial t} &= -\frac{\partial {\cal B}_{z}}{\mu_{0}\partial{x}} - J_{y}\label{eq: elecy}
\end{align}
Here, $z$ is the direction perpendicular to the two-dimensional system, $\epsilon = \epsilon_{0}n^2$ is the local dielectric constant and $\vec{J} = (J_{x}, J_{y})$ are the polarization currents along the $x$ and $y$ directions
\begin{align}
    \vec{J} = \frac{d\vec{P}}{dt} = n_{0}\frac{d(\text{Tr}(\hat{\rho}\vec{\hat{\mu}}))}{dt}
    \label{eq: current}
\end{align}

In eq \ref{eq: current}, $\vec{P}(x, y)$ is the local polarization and $n_{0}$ is the number density of the molecular layer, $\vec{\hat{\mu}} = (\hat{\mu}_{x}, \hat{\mu}_{y})$ are two matrices of transition dipole moments between the molecular states in the $x$ and $y$ directions. 
The molecular layers are modeled as a set of 3-level systems represented by the Hamiltonian:
\begin{align}
\hat{H} = \sum_{a=0}^{2}E_{a}\ketbra{a}{a} + \sum_{a = 1}^{2}V_{a}(t)(\ketbra{0}{a} + \ketbra{a}{0})
\end{align}
Here the molecular subsystem is modeled by a two-dimensional Hydrogen-like atom where the ground state $\ket{0}$ corresponds to the $1s$ orbital and the two excited states $\ket{1}$ and $\ket{2}$ are degenerate, i.e. $E_{1}=E_{2}$, corresponding to the $2p_{x}$, $2p_{y}$ orbitals.\cite{sukharev2011numerical}
The coupling to the EM field within the cavity takes the standard form under dipole approximation:
\begin{align}
    V_{a}(t) = \vec{\mu}_{0a}\cdot \vec{{\cal E}}(t), a=1, 2
\end{align}
Note that the Hamiltonians for all 3-level molecules are explicitly time-dependent, because the electric field $(\vec{{\cal E}(t)}=\bigl({\cal E}_{x}(x,y; t), {\cal E}_{y}(x,y; t)\bigr))$ enters the Hamiltonian in the coupling between the ground state and the doubly-degenerate excited states. 
The spatial position $(x, y)$ is described on a numerical grid, and each grid point in the molecular layer is taken to contain a molecule whose internal 3-state dynamics is described by a $3\times3$ density matrix $\hat{\rho}(t)$. 
These molecular density matrices are propagated in time following the Liouville equation
\begin{align}
    i\hbar\frac{d}{dt}\hat{\rho} &= [\hat{H}(t), \hat{\rho}]
    \label{eq: liouvilleeqmeanfield}
\end{align}
\subsection{Initial condition and the incident pulse\label{subsec: initial condition}}
At time $t=0$, all molecules are assumed to be in the ground state ($\hat{\rho}=\ket{0}\bra{0}$, respectively). Near the edge of the simulation domain (e.g., far outside the cavity), we generate an incident pulse by applying an external EM field along a line of grid points (parallel to the $x$ axis at $y=y_{0}$) with a Gaussian spatial distribution
\begin{align}
{\cal E}_{x}(x, y=y_{0},t) = {\cal E}_{0}\exp{(-x^2/2\sigma_{x}^2)}\times F(t).
\label{eq: shape}
\end{align}
and a temporal pulse with a sine envelope 
\begin{align}
F(t)= 
\begin{cases}
\sin(\pi t/\tau)\sin\omega_{0}t & 0<t<\tau\\
0 & t>\tau
\end{cases}
\label{eq: temporal}
\end{align}
Here, the central frequency of the incident pulse is $\hbar \omega_{0} = 2.41 $eV, the initial spatial size of the external field is $\sigma_{x}=200$, and the pulse duration is $\tau=100$ fs (which determines the pulse shape along $y$ direction). Note that the choices of the spatial size $\sigma_{x}$ and pulse duration $\tau$ do not affect the polariton transport rate.  The incident EM field comprises two components propagating in the $\pm y$ directions respectively. 
The $+y$ component propagates into the cavity and the $-y$ component is absorbed by the perfectly matched layer boundaries (PML) that 
is routinely placed at the edge of the simulation cell.\cite{taflove2005computational} Note that we choose the pulse duration sufficiently long (corresponding to narrow linewidth) so that, when the $+y$ component reaches the cavity, the spatial distribution along the $x$ direction remains focused. After the pulse reach the cavity, the subsequent time evolution includes the molecular excitations followed by polariton propagation and the non-radiative decay. 
For a collection of all parameters and their values in this paper, please see Appendix \ref{apdx: parameters}, Table \ref{tab: para}.

\section{Results \label{sec: three}}
In this section, we present our main results. First, we vary the incident angle (or equivalently, the in-plane wavevector $k_{\parallel}$) and use conventional transfer matrix method (TMM)\cite{yeh1990optical, kavokin2017microcavities} to obtain the dispersion relationships for $(i)$  the bare cavity photon modes (i.e. when the cavity is empty) and $(ii)$ the polaritonic modes (i.e. when the cavity is filled with a complex-valued dielectric). 
Second, after the incident pulse hits the DBR mirror, we investigate how the spatial motion of the EM field is captured by the cavity and how collective transport occurs within the molecular layer inside the cavity. Finally, we compare the group velocities for different $k_{\parallel}$ (or equivalently, $\omega_{0}$) obtained from the dispersion relation of the lower polariton branch and the transport rates obtained by numerically integrating the Maxwell-Bloch equations. 

\subsection{Dispersion relation of the cavity}
We begin by studying the bare cavity photon mode energy as a function of incident angle, i.e. the dispersion relation of the bare cavity modes. 

As is well-known, the velocity of transport of any quasi particle is proportional to the derivative of its dispersion relation line. Moreover, we are aware of that the velocity of light (one component of polariton) is dominated by the dielectric constant of the system. 
Therefore, the geometry of the cavity (as discussed in Section \ref{sec: two} A) and the refractive indices ($n_{1}$ and $n_{2}$) of the two dielectric layers determine the dispersion relation of bare cavity modes. Without loss of generality, we assume $n_{1} < n_{2}$, and we choose $n_{1}=1.005$ (close to air) and $n_{2}=2.52$ (TiO$_{2}$), such that we have very large group velocities, especially at large incident angles. 
Note that we have designed parameters here to give a large cavity mode group velocity, so that the dispersion curve of the lower polariton will be sensitive to the central frequencies of the incident pulse and the Hopfield coefficients;  see Fig \ref{fig: 4} (b) and Fig \ref{fig: 5} (a) below. 
\begin{figure}[ht!]
\includegraphics[width=15cm]{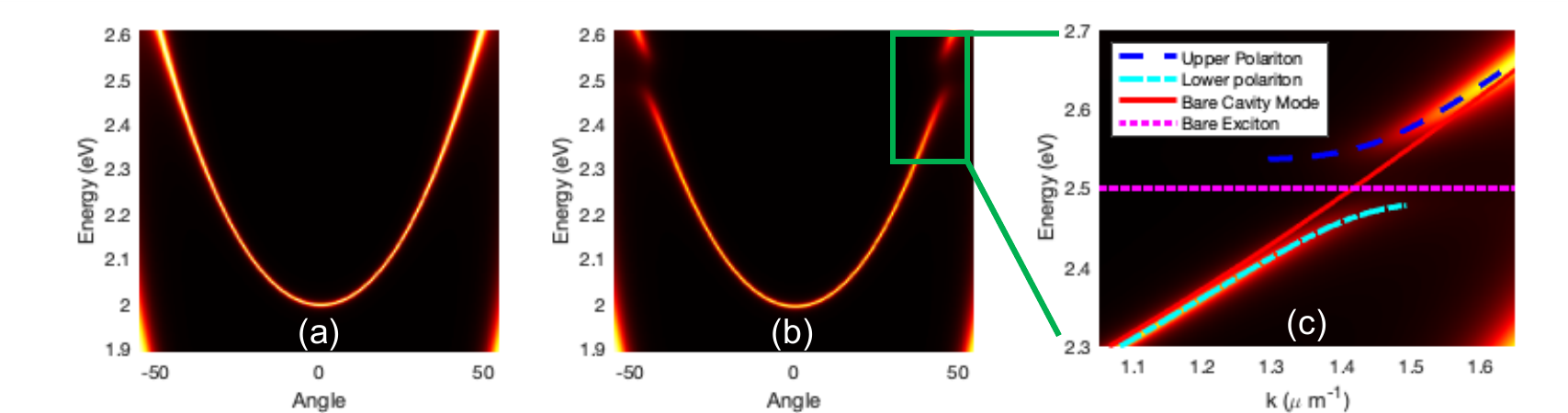}
\caption{Angle-resolved transmission spectrum as a function of energy and incident angles of (a) the empty cavity (without the molecular layer) modes and (b) the polaritons formed (with molecular layer) near $45$ degrees. Both spectra are obtained by transfer matrix method. (c) Magnified dispersion relation (Energy vs wavevector) of polaritons near the avoided crossing at $2.5$ eV. The dispersion lines of polaritons and bare cavity modes are obtained by finding the maximal transmission signal for each $k_{\parallel}$ (see eq \ref{eq: in-plane k}). We also use Maxwell-Bloch equations to obtain the transmission spectrum at $45$ degrees as the incident angle. The transmission spectrum agrees with the results obtained from the transfer matrix method.}
\label{fig: 2}
\end{figure}

Once we have fixed the refractive indices of the two dielectric layers and the central wavelength, angle-resolved transmission spectra can be obtained by applying the standard transfer matrix method (TMM)\cite{yeh1990optical, kavokin2017microcavities} for the system made of two 3-pair distributed Bragg reflectors (see Fig \ref{fig: systemsketch}), as shown in Fig \ref{fig: 2} (a-b). In Fig \ref{fig: 2} (c), we further extract the dispersion relation of lower/upper polaritons ($E_{LP/UP}$) and the bare cavity photonic modes ($E_{cav}$) predicted from the TMM method by numerically finding the maximal transmission value. (The brightest energetic values of the lines on the angular-resolved transmission spectra for each incident angle)
We can extract the group velocity by evaluating the derivative of the energy with respect to the in-plane wavevector ($k_{\parallel}$) along the dispersion relation in Fig \ref{fig: 2} (c):
\begin{align}
v_{\parallel}^{j}=\frac{1}{\hbar}\frac{dE_{j}}{ dk_{\parallel}} \qquad j=\text{cav/LP/UP}
\label{eq: dedk}
\end{align}
Here, $k_{\parallel}$ is defined by
\begin{align}
k_{\parallel}^{j}=\frac{\hbar}{E_{j}c_{0}}\sin{\theta} \qquad j=\text{cav/LP/UP}
\label{eq: in-plane k}
\end{align}
where $\theta$ is the incident angle and $c_{0}$ is the speed of light in vacuum.

Note that in Fig \ref{fig: 2} (a-b),  for normal incidence ($k_{\parallel}\approx 0$), the group velocity of bare cavity photon mode ($E_{cav} = 2eV$) is nearly $0$.
In other words, as one would expect, when the light beam enters the cavity at the normal incidence, the refracted and reflected beams remain completely perpendicular to the cavity dielectric layers and the light does not propagate along an in-plane direction.

To observe polaritons transport, we must apply the EM field at a finite incident angle (chosen to be $45$ degrees for numerical simplicity). Thus, we choose the molecular excitation energy to be $E_{2}-E_{0} = E_{1}-E_{0} = 2.5\text{ eV}$. Following this parameter choice, we can obtain the dispersion line with TMM using the dielectric constant (see eq 41 in ref\citenum{sukharev2017optics}, it is the Drude model for a two-level system).
As shown in Fig \ref{fig: 2}(b), the avoided crossings confirm that polaritons are indeed formed near $2.5\text{ eV}$. 
The transmission spectrum (with two polariton peaks) at $45$ degrees incidence has also been verified using Maxwell-Bloch equations with a short-time light pulse (same functional form as in eqs \ref{eq: shape} and \ref{eq: temporal}, but with $\tau=10 \text{ fs}$).
These dispersion lines are later used to determine the light-matter interaction and the Hopfield coefficients (see Fig \ref{fig: 5} and Section \ref{sec: discussion}).

Finally, as a side note, we want to remind the reader of one limitation of our light source. 
In theory, the molecular layer with excitation energy $2.5$ eV is  strictly resonant with only the $45$ degree bare cavity mode. 
However,  the molecular layer does formally interact with a continuum of cavity modes that are close to $2.5$ eV. 
Moreover, because the incident light is spatially localized (in contrast to completely coherent and parallel light source), the envelope of the incident light will spread along the $x$ direction when traveling towards the positive $y$ direction. 
Thus, in our FDTD simulation, the effective incident angle has a natural spread around $45$ degrees. 
Below, we will use this imperfection of the light source to our advantage, insofar as the spread in angles will allow us to pump the lower polariton states (which have a spread in energy between 2.41 and 2.47 eV) by simply varying the center frequency $\omega_{0}$ (as in eq \ref{eq: shape}) and not worrying about matching wavevectors (i.e. there is no need to change the incident angle).

\begin{figure}[ht!]
\includegraphics[width=15cm]{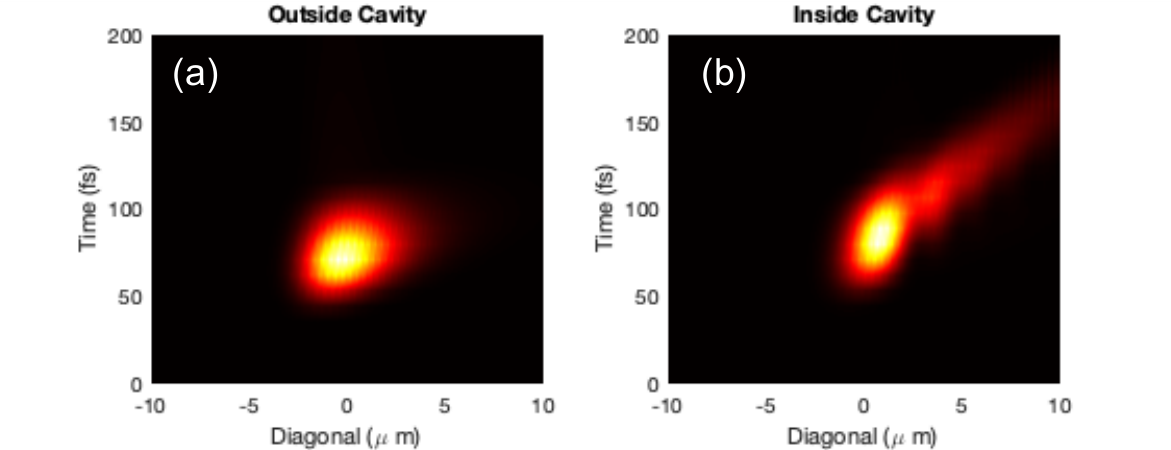}
\caption{Excited state population of (a) bare molecular layer and (b) molecular layer inside a DBR cavity as in Fig \ref{fig: systemsketch}, plotted as a function of time and position along the direction of the cavity mirror/molecular layer. For the case outside the cavity, because the molecular layer is along the diagonal of the simulation box while the incoming light field is moving along $y$ direction, the excited state population is not symmetric with respect to the diagonal direction. Although the excitation spot slightly leans towards right side, the population map does not show obvious transport behavior. For the case inside the cavity in (b), we can clearly observe that after the pulse create an excitation spot near the center, the excited state population is transported in time to the right side of the box, as time ($y$ axis) flies.}
\label{fig: 3}
\end{figure}
\subsection{Polariton transport}
Having confirmed the existence of polaritonic states, we will now proceed to study the polaritonic transport. In the context of a Maxwell-Bloch treatment, because the EM field is treated in the position (and not mode) representation , we cannot easily compute the population of the cavity mode. 
Instead, we focus on the evolution of the molecular excitations in real space, assuming that at any point in space, the molecular excitation synchronise with the local EM field intensity. This serves as a way to effectively capture the polariton propagation.
Fig \ref{fig: 3} shows the evolution of the molecular excitation density following excitation by a nearly monochromatic pulse centered near lower polariton resonance ($\hbar \omega_{0} = 2.41 $eV) of duration $100$ fs.
Both figures in Fig \ref{fig: 3} show a heat map of molecular excitation (which starts as a spot). Outside the cavity, no tail appears around the spot. 
In contrast, as shown in Fig \ref{fig: 3}(b), for the case inside a cavity, after the excited state population reaches a maximum near $50$ fs, the spot moves towards positive $x$ direction and shows a long tail.
We can extract the transport rate of polaritons by numerically calculating the slope of this tail.

\subsection{Comparing the TMM Group Velocity and the Maxwell-Bloch Transport Rates}
\begin{figure}[ht!]
\includegraphics[width=15cm]{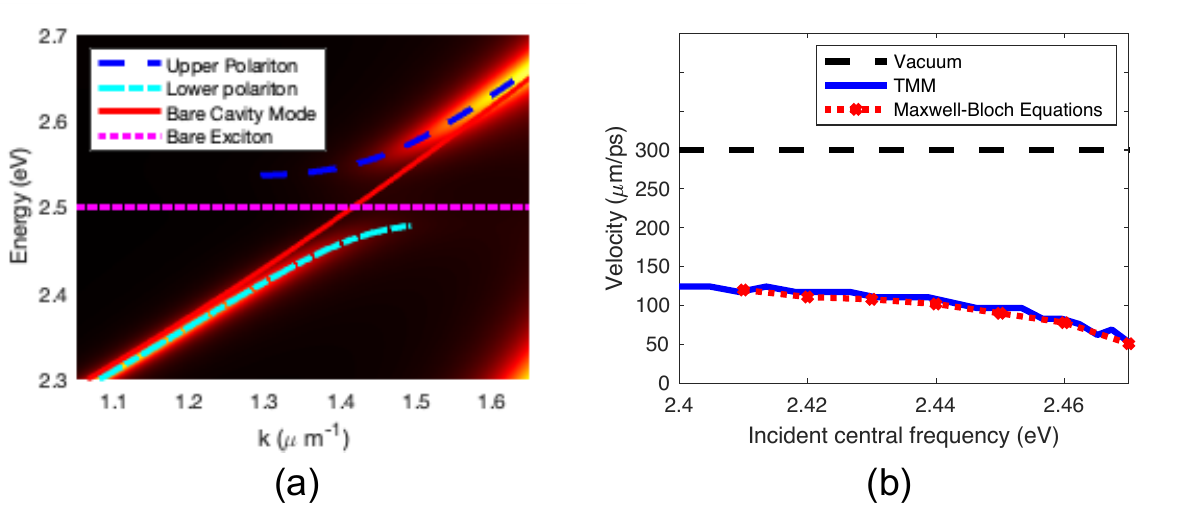}
\caption{(a) The dispersion relation of the lower (cyan dashed-dotted line) and upper (blue dash line) polaritons, bare exciton (magenta dotted line) and bare cavity photon modes (red solid line). (b)Observed transport velocity calculated from the dispersion relation (blue solid line) and Maxwell-Bloch calculations (red dotted line with crosses) as a function of central frequency of the incoming pulse. The vacuum speed of light (black dashed line, $300 \mu \text{m/ps}$) is much larger than the polariton transport velocities inside the cavity. The dispersion relation group velocity is obtained according to eq \ref{eq: dedk} and by calculating the first derivative of the lower polaritons dispersion line. }
\label{fig: 4}
\end{figure}

Fig \ref{fig: 4}(a) reproduces the results of Fig. \ref{fig: 2}(c) above. 
When the polaritons are formed, the avoided crossing near the molecular resonance strongly distorts the dispersion lines, and the group velocities of the lower and upper polaritons are drastically changed from the original group velocities of the bare cavity photon (as explicitly expressed by eq \ref{eq: apdx10} below). 
In blue in Fig \ref{fig: 4}(b), we plot the corresponding group velocity associated with the lower polariton as calculated in the energy range of $2.41-2.47$ eV from eq \ref{eq: dedk} using the dispersion obtained from the TMM calculation. The value compared well with the speed (red dotted line) calculated from the slope of the propagation tail seen in Fig \ref{fig: 3}(b) that was extracted from the Maxwell-Bloch simulations. The small discrepancy may arise from the imperfect light source. 
The calculated speed is of the order of $1/3$ of the speed of light in vacuum (dashed black line), but still represents a considerable speed for propagation involving molecular excitation. We note that the group velocity of the cavity photon, as calculated from the cavity photon dispersion is approximately $160 \mu \text{m/ps}$. While the calculated number is smaller than that reported in Ref \citenum{balasubrahmaniyam2023enhanced}, we note that the speed is nearly the largest we could get with this combination of cavity geometry and excitation process, by using $n_{1}=1.005\approx 1.00$. If we choose more realistic parameters, for example, $n_{1} = 1.48$ as silica, the group velocities will become much smaller. Finally, we briefly discuss the effects of $n_{2}$ (quality factor) and show the corresponding numerical results in Appendix \ref{apdx: 3}.
\section{Discussion\label{sec: discussion}}
The polariton states are considered as linear combinations of molecular excitation and cavity photon states.
\begin{align}
\ket{LP/UP} = \alpha_{ex}^{LP/UP}\ket{ex} + \alpha_{cav}^{LP/UP} \ket{cav}
\label{eq: lpstate}
\end{align}
A simple consideration of the dispersion relation that would be obtained from these equations leads to the group velocities and their relation to the Hopfield coefficients for varying incident angles, or equivalently, in-plane wavevectors ($k_{\parallel}$). Explicitly, 
\begin{align}
    v^{LP/UP}_{g}=\vert\alpha_{cav}^{LP/UP}\vert^2 v^{cav}_{g} \pm \frac{d\vert V\vert}{dk_{\parallel}}\vert\alpha_{ex}^{LP/UP}\alpha_{cav}^{LP/UP}\vert
    \label{eq: apdx10}
\end{align}
The full derivation is presented in Appendix \ref{apdx: 0}. The first term is intuitively easy to understand, the group velocities of the lower and upper polaritons depend on their portion of photonic components. 
In other words, the more photonic component a polariton state contains, the faster the group velocity is. 
For instance, when the bare exciton and the bare cavity photon mode are resonant at $k_{\parallel}\approx 1.415 \mu m^{-1}$ (incident angle $\theta = 45$ degrees), both polaritons possess $50\%$ photon component and thus, are expected to transport with the same velocity, which is that of a half of the group velocity of bare cavity photon mode. Moreover, when the energy of bare cavity photon mode is very close to that of the lower polariton, i.e., $k_{\parallel} < 1.35 \mu m^{-1}$, the group velocity of the lower polariton is approximately the same as that of the bare cavity photon, and the group velocity of the upper polariton is approximately $0$ (as shown in Fig \ref{fig: 4}(a), the cyan line is very close to the red line, and the blue line is very close to the magenta line, respectively). More generally, the fraction of the photon/exciton occupation in the polariton state is determined by the Hopfield coefficients $\vert\alpha_{cav/ex}\vert^2$. 

\begin{figure}[ht!]
\includegraphics[width=15cm]{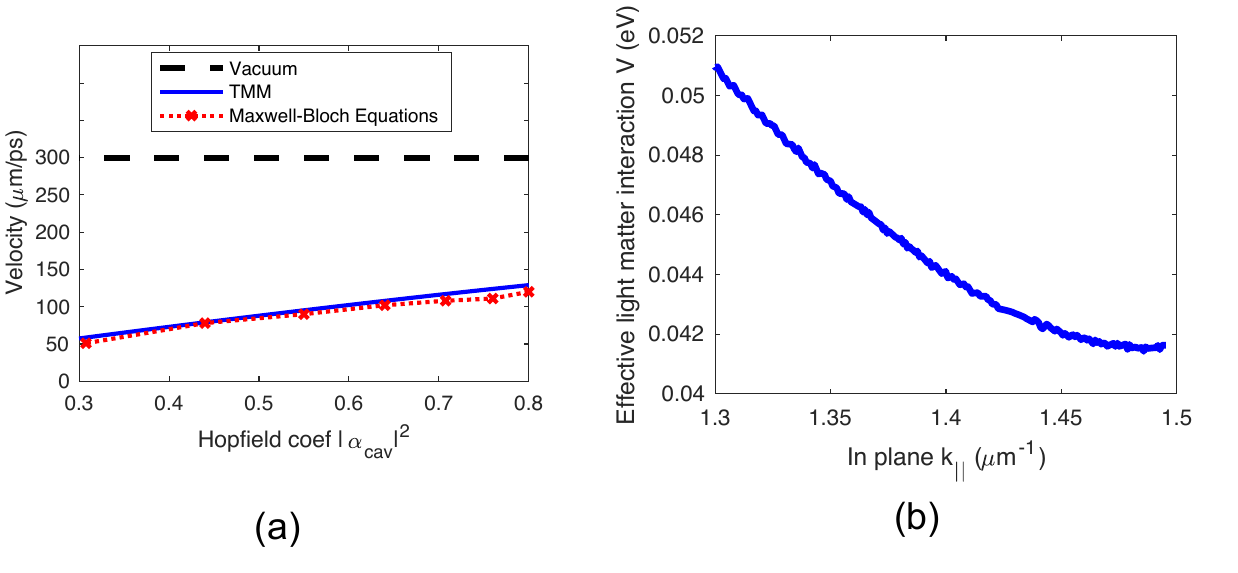}
\caption{(a)Observed transport velocity from the dispersion relation (blue solid line) and Maxwell-Bloch calculations (red dotted line with crosses) as a function of Hopfield coefficients of lower polaritons and (b) effective light-matter interaction $\vert V(k_{\parallel})\vert$ obtained from TMM calculations. }
\label{fig: 5}
\end{figure}

If we assume that the magnitude of light-matter interaction $\vert V\vert$ is weakly dependent on the in-plane wavevector $k_{\parallel}$ (in contrast to the cavity mode energy $E_{cav}$) so that the second term is small, the first term agrees with the data previously presented:  the transport rate is primarily dominated by the bare cavity photon transport rate ($v^{cav}_{g}$) and the Hopfield coefficient $\vert\alpha_{cav}^{LP/UP}\vert^2$. 
This allows for a simple classical picture of the polariton transport. 
Namely, for the initial near-monochromatic excitation that matches a given polariton energy, the excitation induced in the system appears partly in the EM field inside the cavity and partly in molecular excitation, divided according to their corresponding Hopfield coefficients. 
In the strong coupling regime, the two portions reaches equilibrium much faster than the effective transport and cavity leakage rates, hence the observed rate is a weighted average of the group velocity of the cavity photon modes (which is solely governed by the mirror properties). 
The molecular excitation effectively does not move on the timescale of observation. 
Indeed, in Fig \ref{fig: 5}(a), we plot polariton speed as a function of $\vert\alpha_{cav}\vert^2$. We find that the result is basically a straight line, because within the very small energy window $2.41-2.47$ eV, the group velocity of the bare photon does not significantly change.
Note that the second term on the right side of eq \ref{eq: apdx10} is a product of the derivative of the light-matter coupling term ($V$) with respect to the in-plane wavevector $k_{\parallel}$ multiplied by  the coherence between the cavity mode and the quantum molecular subsystem. The magnitude of $V$ can be estimated from a TMM calculation by insisting that the eigenvalues of the matrix in eq \ref{eq: diaglp} (see Appendix \ref{apdx: 0}) are consistent with the values of $E_{cav}, E_{ex}, E_{LP}$ and $E_{UP}$ in Fig. \ref{fig: 4}(a). 
In Fig \ref{fig: 5} (b), we plot $\vert V \vert$ as a function of $k_{\parallel}$. Clearly, $\vert V \vert$ is not a constant. However, the derivative of $V$ is less than $1/10$ of the group velocity of the bare cavity photon and so this term appears to be small. 

This analysis has important implications for understanding experiments. In many cases, the assumption that the light-matter coupling $\vert V\vert$  is constant -- or equivalently, that the Rabi splitting $\Omega_{R}$ is constant --  with respect to small incident angle/energy changes is a valid one. However, in the context of polariton transport, this assumption obscures the quantum nature of polaritons. In principle, polaritons need not function as simple averages of the cavity photon mode and molecular excited states, especially in the strong collective coupling regime ($\vert V\vert$ being large),
and we hope that the present analysis will trigger further investigations on polariton transport where both terms in eq \ref{eq: apdx10} contribute to the transport.

\section{Conclusion and Outlook\label{sec: four}}
In conclusion, we have performed calculations based on a Maxwell-Bloch formalism within a mean field approximation to simulate polariton transport phenomena for a two dimensional system.
The polariton transport is represented as the synchronized motion of molecular excitations and the EM field inside a cavity.
The transport rate is in excellent agreement with the group velocity as obtained from the transfer matrix method (and we attribute the small errors to the finite system size and a non-ideal light source).
Moreover, the Maxwell-Bloch equations formalism gives us a very intuitive picture of  polariton transport, as one can directly monitor the such transport by quantifying either the molecular excitations or the EM field intensities in the real space.
In the present case, we find that the transport has a very simple functional form, whereby the velocity is simply the hopfield coefficient multiplied by the speed of light in a raw cavity.
As a side note, we mention that the calculated timescale  for polariton transport phenomena is less than $100 fs$, the total length of our simulation; this timescale is much faster than the timescale ($\approx1 ns$) for F\"{o}rster resonance energy transfer (FRET) between a donor and acceptor, even when enhanced by a  cavity\cite{andrew2000forster}. 

Looking forward, one can directly introduce different types of disorder to reduce the polariton transport rate and hopefully observe simultaneously ballistic transport and diffusive behaviors. 
In this work, when working with two dimensional calculations, we could focus only on the transverse magnetic (TM) cavity modes. In the future, though it would be more computationally expensive, one could perform the calculation also on a three dimensional system and observe the interplay between transverse electric (TE) and transverse  magnetic (TM), where the results would be most interesting if the molecules had non-trivial polarizability tensors and were able to change light polarization. Extending the present work to more complicated materials, with more geometric degrees of freedom, will be an important next step for  
further understanding cavity-promoted polariton transport phenomena from first principles. 

\section*{Acknowledgements}
This work has been supported by the U.S. Department of
Energy, Office of Science, Office of Basic Energy Sciences,
under Award No.DE-SC0019397 (J.E.S.); the U.S. National
Science Foundation under Grant No.CHE1953701 (A.N.). MS acknowledges support by the Air Force Office of Scientific Research under Grant No. FA9550-22-1-0175.

\appendix

\section{Effects of quality factor \label{apdx: 3}}
As discussed in section \ref{sec: three}, the quality factor of the cavity is governed by the larger refractive index ($n_{2}$). This fact can be easily understood by noting that the reflection coefficient near resonant mode at normal incidence of a $N$-pair DBR mirror with refractive indices $n_{1}$ and $n_{2}$ is
\begin{align}
    r_{DBR} = \frac{n_{2}^{2N}-n_{1}^{2N}}{n_{1}^{2N}+n_{2}^{2N}} = 1-\frac{2n_{1}^{2N}}{n_{1}^{2N}+n_{2}^{2N}}
    \label{eq: refcoef}
\end{align}
Thus, the EM field trapped inside the cavity becomes more difficult to transmit the mirrors. In other word, the cavity generates a larger field-amplification factor and a longer lifetime. For a detailed derivation of eq \ref{eq: refcoef} and an overview of the standard transfer matrix method for any incident angle, there are many great textbooks on optics, for example, see ref \citenum{yeh1990optical, kavokin2017microcavities}.

\begin{figure}[ht!]
\includegraphics[width=12cm]{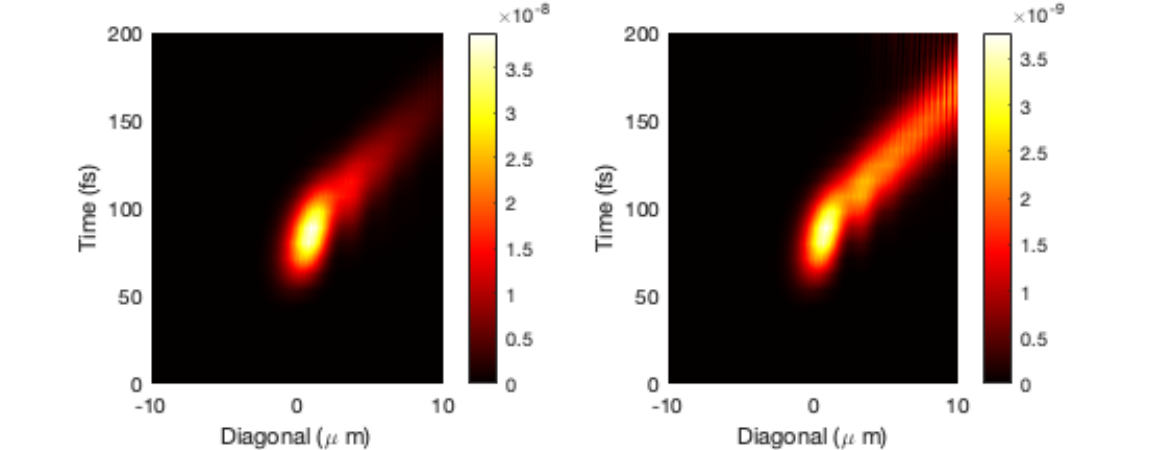}
\caption{Excited state population of molecular layer inside a cavity with (a) small $n_{2} = 2.5$ and (b) large $n_{2} = 7.5$ contrast between dielectric layers. Fig \ref{fig: 6}(a) is identical to Fig \ref{fig: 3}(b) but with a colorbar showing the actual populations. Obviously the transport rates extracted from the two scenarios are identical. However, the signal in Fig \ref{fig: 6}(b) lasts longer.}
\label{fig: 6}
\end{figure}
We can minimally distort our simulation by simply increase $n_{2}$ and decrease layer thickness $L_{2}$ according to eq \ref{eq: layerthicknesses}. Again, this alteration does not affect the slope of the dispersion relation and therefore, does not affect the transport rate observed from Maxwell-Bloch calculations. As shown in Fig \ref{fig: 6}, on the one hand, as the quality factor increases from scenario (a) to (b), the signal intensity decreases ($4$ times smaller), because the EM field enters the cavity with less efficiency. On the other hand, once the EM field enters the cavity, it remains trapped and transport for a longer distance. Therefore, refining quality factor increases the fidelity of the signals.
\section{Derivation of the relation between polariton group velocities and bare cavity mode group velocities \label{apdx: 0}}
In this appendix, we present the detailed derivation in eq \ref{eq: apdx10}
For a system with one bare excitation exciton energy ($E_{ex}$), cavity mode energy ($E_{cav}$), and an approximate collective light-matter coupling $V = \Omega_{R}/2$ ($\Omega_{R}$ is the Rabi splitting), the standard $2\times 2$ Hamiltonian is:\begin{align}
\hat{H}&=  \left[
\begin{array}{c c}
E_{ex} & V\\ 
V^{*}& E_{cav} \\ 
\end{array} \right]
\label{eq: hamforhopfield}
\end{align}
 The lower and upper polaritons energies $E_{LP/UP}$ 
are the eigenvalues of this Hamiltonian,
 with eigenvectors $\ket{LP}, \ket{UP}$. Let us focus on the lower polariton state $\ket{LP}$
\begin{align}
    \left[
\begin{array}{c c}
E_{ex} & V\\ 
V^{*}& E_{cav} \\ 
\end{array} \right]
\left[
\begin{array}{c}
\alpha_{ex}^{LP}\\ 
\alpha_{cav}^{LP}\\ 
\end{array} \right]
&=
E_{LP}\left[
\begin{array}{c}
\alpha_{ex}^{LP}\\ 
\alpha_{cav}^{LP}\\ 
\end{array} \right]
\label{eq: diaglp}
\end{align}
which clearly satisfies:
\begin{align}
    \frac{\alpha_{ex}^{LP}}{\alpha_{cav}^{LP}}=\frac{V}{E_{LP}-E_{ex}} = \frac{E_{LP}-E_{cav}}{V^{*}}
    \label{eq: relationbetweenalphaandv}
\end{align}
Clearly, both $\frac{\alpha_{ex}^{LP}}{V\alpha_{cav}^{LP}}$ and $\frac{V^{*}\alpha_{ex}^{LP}}{\alpha_{cav}^{LP}}$ are real-valued.
Now, the standard characteristic equation is:
\begin{align}
    (E_{LP}-E_{cav})(E_{LP}-E_{ex}) -\vert V\vert^2=0
    \label{eq: characteristic_eq}
\end{align}
If we take derivative of this equation with respect to $k_{\parallel}$, we obtain
\begin{align}
    (\frac{dE_{LP}}{dk_{\parallel}}-\frac{dE_{cav}}{dk_{\parallel}})(E_{LP}-E_{ex}) + (\frac{dE_{LP}}{dk_{\parallel}})(E_{LP}-E_{cav})-V^{*}\frac{dV}{dk_{\parallel}} - V\frac{dV^{*}}{dk_{\parallel}}=0
    \label{eq: firstder17}
\end{align}
Here, we naturally assume that the exciton energy is independent of the in-plane wavevector $k_{\parallel}$ of the incident photon. Therefore, we can combine eqs \ref{eq: firstder17} and \ref{eq: relationbetweenalphaandv} to obtain:
\begin{align}
    (\frac{dE_{LP}}{dk_{\parallel}}-\frac{dE_{cav}}{dk_{\parallel}})(\frac{\alpha_{cav}^{LP}}{\alpha_{ex}^{LP}}V + \frac{\alpha_{cav}^{LP*}}{\alpha_{ex}^{LP*}}V^{*})/2 + (\frac{dE_{LP}}{dk_{\parallel}})(\frac{\alpha_{ex}^{LP}}{\alpha_{cav}^{LP}}V^{*} + \frac{\alpha_{ex}^{LP*}}{\alpha_{cav}^{LP*}}V)/2
    \\
    \nonumber
    -V^{*}\frac{dV}{dk_{\parallel}} - V\frac{dV^{*}}{dk_{\parallel}}=0
\end{align}
which simplifies to (using $\vert\alpha_{ex}^{LP}\vert^2 + \vert\alpha_{cav}^{LP}\vert^2 = 1$):
\begin{align}
    \frac{dE_{LP}}{dk_{\parallel}}(\frac{V}{\alpha_{cav}^{LP*}\alpha_{ex}^{LP}} + \frac{V^{*}}{\alpha_{cav}^{LP}\alpha_{ex}^{LP*}})-\frac{dE_{cav}}{dk_{\parallel}}(\frac{\alpha_{cav}^{LP}}{\alpha_{ex}^{LP}}V + \frac{\alpha_{cav}^{LP*}}{\alpha_{ex}^{LP*}}V^{*}) =2V^{*}\frac{dV}{dk_{\parallel}} + 2V\frac{dV^{*}}{dk_{\parallel}}
\end{align}
or even better,
\begin{align}
    \frac{dE_{LP}}{dk_{\parallel}}-\vert\alpha_{cav}^{LP}\vert^2\frac{dE_{cav}}{dk_{\parallel}}=\frac{\frac{d\vert V\vert^2}{dk_{\parallel}}}{\frac{E_{LP}-E_{ex}}{\vert\alpha_{cav}^{LP}\vert^2} + \frac{E_{LP}-E_{cav}}{\vert\alpha_{ex}^{LP}\vert^2}}=\frac{\frac{d\vert V\vert^2}{dk_{\parallel}}}{\frac{V}{\alpha_{cav}^{LP*}\alpha_{ex}^{LP}} + \frac{V^{*}}{\alpha_{ex}^{LP*}\alpha_{cav}^{LP}}}
    \label{eq: apdx9}
\end{align}
Note that, without loss of generality, one is free to choose the overall phase of the $\ket{LP}$ eigenvector (eq \ref{eq: lpstate}), such that we can assume that $\alpha_{ex}^{LP} = \alpha_{ex}^{LP*}$ is real-valued. 
From eq \ref{eq: relationbetweenalphaandv}, it then follows that $V/\alpha_{cav}^{LP*} = V^{*}/\alpha_{cav}^{LP} =\vert V\vert/\vert\alpha_{cav}^{LP*}\vert$ is also real-valued.
And so, in the end, we can further simplify eq \ref{eq: apdx9} according to eq \ref{eq: relationbetweenalphaandv}
\begin{align}
    \frac{dE_{LP}}{dk_{\parallel}}=\vert\alpha_{cav}^{LP}\vert^2\frac{dE_{cav}}{dk_{\parallel}} + \frac{d\vert V\vert}{dk_{\parallel}}\vert\alpha_{ex}^{LP}\alpha_{cav}^{LP}\vert
    \label{eq: apdx11}
\end{align}
\section{Parameters for Maxwell-Bloch simulation\label{apdx: parameters}}
In this appendix, we list all parameters in our simulation.
\begin{table}[ht!]
  \begin{threeparttable}
   \caption[]{Parameters for Maxwell-Bloch simulation.}
   \centering
   \label{tab: para}
   \begin{tabular}{P{8cm}c}
     \midrule 
     Name & Value\tnote{$\dagger$}
    \\\hline
    Grid resolution ($dx$) & $4$ nm
    \\
    Time step (dt) & $dx/1.5/c_{0}$
    \\
    Thickness of Molecular layer & $40/\sqrt{2}$ nm
    \\
    Number of pairs of dielectric materials & $3$ pairs
    \\
    Thickness of layer 1& $154$ nm 
    \\
    Thickness of layer 2& $248$ nm 
    \\
    Distance between inner surface of the two mirrors & $156$ nm
    \\
    Transition dipole moment & $\mu_{x}^{01} = \mu_{y}^{02}= 10/\sqrt{3}$ Debye
    \\
    Pulse duration & $\tau=100$ fs \\
    \midrule
     \end{tabular}
  \end{threeparttable}
\end{table}
\newpage
\bibliography{apssamp}

\end{document}